\documentclass[a4paper,10pt,hidelinks]{article}
\usepackage[utf8]{inputenc}
\usepackage[T1]{fontenc}
\usepackage{units}
\usepackage{siunitx}
\usepackage{url}
\usepackage{hyperref}
\usepackage{amsmath}
\usepackage{amssymb}
\usepackage{graphicx}
\usepackage[dvipsnames]{xcolor}
\usepackage[version=4]{mhchem}

\usepackage{stfloats} %

\usepackage[labelfont=bf, justification=justified, font=footnotesize]{caption}

\usepackage[style=numeric-comp,maxnames=25,sorting=none,doi=false,date=year,isbn=false,firstinits=true]{biblatex}

\renewbibmacro{in:}{}
\DeclareFieldFormat*{url}{}
\DeclareFieldFormat[report]{url}{\mkbibacro{URL}\addcolon\space\url{#1}}

\usepackage[a4paper, twocolumn, left=2.0cm, right=2.0cm, bottom=3cm, top=2.8cm, columnsep=0.75cm]{geometry}

\usepackage[symbol]{footmisc}

\usepackage{authblk}

\usepackage{tikz}           %
\usetikzlibrary{calc,decorations.pathmorphing,decorations.pathreplacing,fadings,shadings,shapes,arrows,positioning,patterns,automata,shadows.blur,fit,arrows.meta}

\usetikzlibrary{arrows.meta}

\DeclareRobustCommand\filledleftrightarr{\begin{tikzpicture}\draw [{Latex}-{Latex}](0,0) -- (0.5,0);\end{tikzpicture}}
\DeclareRobustCommand\emissionarr{\begin{tikzpicture}\draw [{Circle[open]}-{Triangle[open]}](0,0) -- (0.5,0);\end{tikzpicture}}

\onecolumn
\date{}

\author[1,2]{Tim Hempel}
\author[3,*]{Simon Olsson}
\author[1,2,4,5,$\dagger$]{Frank No\'e}
\affil[1]{Freie Universit\"at Berlin, Department of Mathematics and Computer Science, Berlin, Germany}
\affil[2]{Freie Universit\"at Berlin, Department of Physics, Berlin, Germany}
\affil[3]{Chalmers University of Technology, Department of Computer Science and Engineering, Göteborg, Sweden}
\affil[4]{Rice University, Department of Chemistry, Houston, TX, USA}
\affil[5]{Microsoft Research, Station Road, Cambridge, United Kingdom}

\title{Markov Field Models: scaling molecular kinetics approaches to large molecular machines}

\begin{document}
\sloppy

\twocolumn[ 
  \begin{@twocolumnfalse}
    \maketitle

\vspace{-1.2cm}

\begin{abstract}

With recent advances in structural biology, including experimental techniques and deep learning-enabled high-precision structure predictions, molecular dynamics methods that scale up to large biomolecular systems are required. Current state-of-the-art approaches in molecular dynamics modeling focus on encoding global configurations of molecular systems as distinct states. This paradigm commands us to map out all possible structures and sample transitions between them, a task that becomes impossible for large-scale systems such as biomolecular complexes.
To arrive at scalable molecular models, we suggest moving away from global state descriptions to a set of coupled models that each describe the dynamics of local domains or sites of the molecular system.
We describe limitations in the current state-of-the-art global-state Markovian modeling approaches and then introduce Markov Field Models as an umbrella term that includes models from various scientific communities, including Independent Markov Decomposition, Ising and Potts Models, and (Dynamic) Graphical Models, and evaluate their use for computational molecular biology. Finally, we give a few examples of early adoptions of these ideas for modeling molecular kinetics and thermodynamics.

\vspace{0.3cm}

\end{abstract}
  \end{@twocolumnfalse}
]
\footnotetext{\hspace{-5mm}\textbf{Corresponding authors:}\\ * simonols@chalmers.se, $\dagger$ frank.noe@fu-berlin.de\\
\textbf{Declaration of interest:} none}

\section*{Introduction}

Computer simulations such as molecular dynamics (MD) are established tools for understanding the function of molecular machines on an atomistic scale. 
In contrast to experiments, \textit{in silico} methods are not limited by their spatial or temporal resolution; their Achilles' heel is that enough data must be gathered to describe a biological system in thermodynamic equilibrium. 
Many recent advances have contributed to the solution of this so-called sampling problem, such as hardware developments like fast graphical processing units (GPUs), efficient software packages \cite{eastman_openmm_2017,abraham_gromacs_2015}, and enhanced sampling methods that, e.g., use bias potentials along a reaction coordinate \cite{laio_escaping_2002,valsson_enhancing_2016,tiwary_spectral_2016,tiwary_metadynamics_2013,ribeiro_reweighted_2018,bonati_deep_2021} or diffusion maps \cite{preto_fast_2014}.
Additionally, Markov State Models (MSMs) have leveraged fast parallel processing power by combining large numbers of short off-equilibrium trajectories without defining reaction coordinates or introducing bias potentials to the system \cite{schutte_direct_1999,swope_describing_2004,singhal_using_2004,chodera_automatic_2007,noe_constructing_2009,prinz_markov_2011}. MSMs have been profiting substantially from the development of deep learning methods in recent years \cite{mardt_vampnets_2018,chen_nonlinear_2019,sidky_highresolution_2019}; see Ref.~\cite{noe_machine_2020} for an overview of both shallow and deep machine learning (ML) methods in this area.

These combined efforts have been very successful, shedding light on complex molecular processes such as protein folding \cite{lindorff-larsen_how_2011,noe_hierarchical_2007,voelz_molecular_2010,voelz_slow_2012}, ligand-protein binding \cite{silva_role_2011,sengupta_markov_2018,plattner_protein_2015,buch_complete_2011}, or even protein-protein association kinetics \cite{plattner_complete_2017}. These small to medium-sized protein systems are often cooperative, giving rise to a small number of rare-event processes between a few long-lived, metastable states, and thus MSMs or other kinetic models can be used to characterize their rare-event dynamics globally \cite{olsson_dynamic_2019}.

However, this approach does not scale with increasing size of the molecular system, as its cooperativity decreases and thus the number of globally distinct metastable states increases combinatorially. As an example, consider a solution of $N$ dissociated proteins \cite{mardt_deep_2022}. If each protein can be in one of two states, the number of all global states is $2^N$, i.e., it scales exponentially with the number of constituents. Therefore, sampling each global system configuration and the transitions between them becomes infeasible even for a small number of proteins. A biomolecular complex, of course, has more cooperativity, and the coupling between domains may reduce the number of globally accessible states. However, the fundamental problem remains, as an increasing number of loosely coupled domains will lead to an exponentially increasing number of global system states. Therefore, even though all-atom MD simulations can now be conducted with impressive system sizes such as a virus in an aerosol particle \cite{dommer_covidisairborne_2021} or a membrane model of the endoplasmic reticulum \cite{trebesch_embracing_2020}, they will not lead to the ability to directly parameterize a global state model (e.g., an MSM) in the near future. This task would require us to increase the aggregate simulation time exponentially with system size, whereas it typically decreases with system size in reality.

\begin{figure}
	\centering
	\includegraphics[width=\columnwidth]{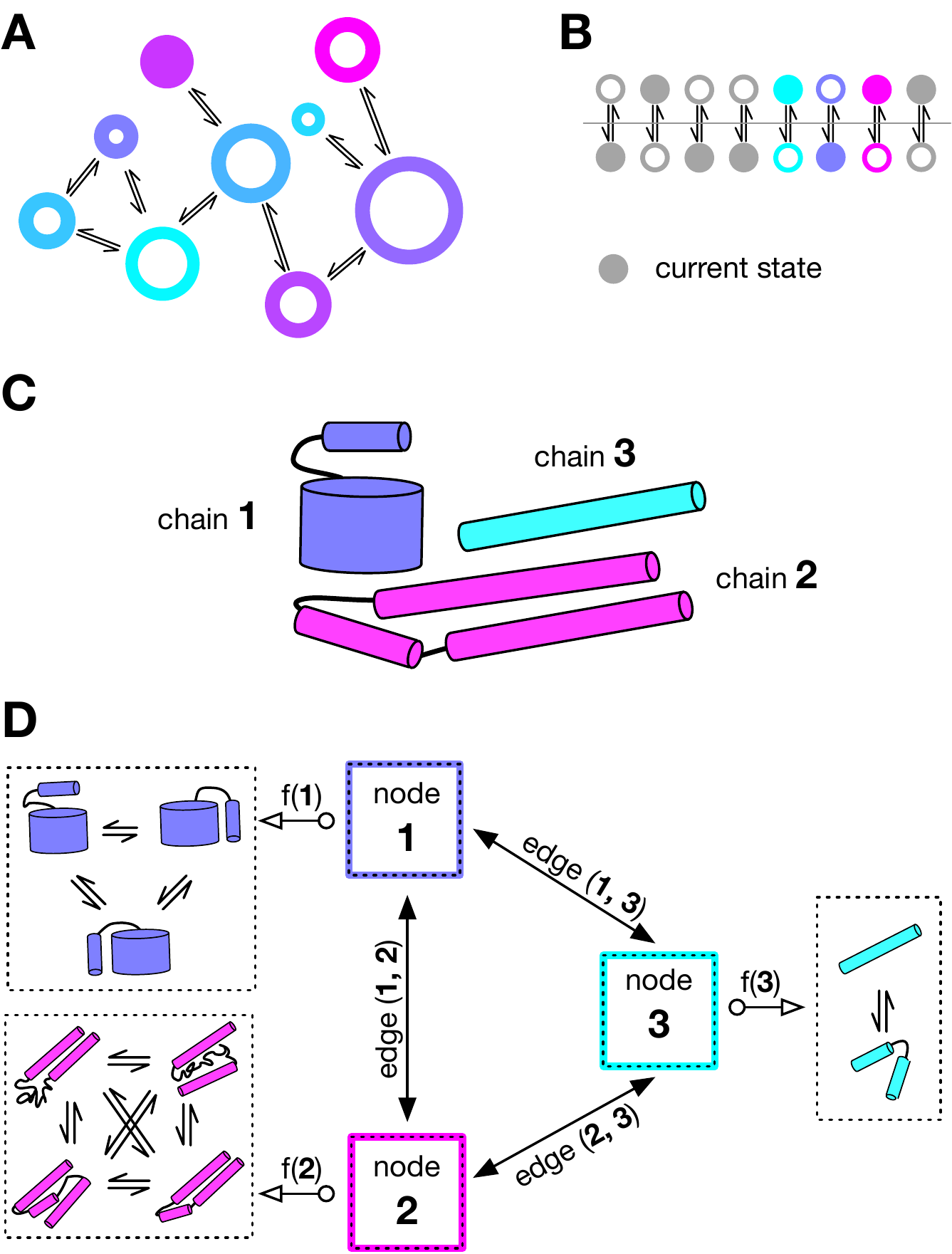}
	\caption{\textbf{Markov field models: From a biomolecular complex to a graph of stochastically coupled subdomains.}
	\textbf{a:} MSM with $M$ distinct, global states that have different equilibrium probabilities (denoted by circle size). Arrows indicate where direct transitions between states are possible. The instantaneous state of the system is defined by being in a single state, here indicated by the filled circle.
	\textbf{b:} Ising model with $N$ local spins. Each Ising spin can be in one of two states. The instantaneous state of the system is defined by the combination of all spin settings, here indicated by the filled circles. Three spins, corresponding to three MFM domains, are highlighted by color-coding as in subsequent panels.
	\textbf{c:} Example of a molecular complex, where each of three protein chains is modeled as a dynamical subsystem. \textbf{d:}  Graph view of the molecular complex, each protein chain is now seen as a node in a graph. The edges (\filledleftrightarr) denote interaction terms, i.e., where the state transitions of each domain can be coupled. Next to each node (\emissionarr), the states and transitions between states of that node or protein subsystem are shown.
	}
	\label{fig:graph}
\end{figure}

The main idea proposed in this manuscript is to avoid the exponential scaling by adopting ideas from Ising models (Fig.~\ref{fig:graph}a,b), a multiscale approach that explicitly recognizes that there are loosely coupled subsets of the complex structure (``domains'', loosely associated with ``spins'' in an Ising model).
Instead of modeling a biomolecular complex as a global entity, we propose to describe its dynamics by a graph consisting of domains (e.g., protein domains) that interact via edges (i.e., coupling) \cite{harary_dynamic_1997}, thus forming the full sequence protein complex (or assembly of domains) as shown in Fig. \ref{fig:graph}c,d.
In this setting, each domain has a limited number of states that can be sampled, and the coupling depends only on local states (e.g. of the pair of coupled states) without the need to explicitly encode the global state. Like in an Ising model, the global dynamics arises from a combination of simple parts.

Although Ising models are established in statistical physics, marrying them with kinetic models of biomolecular complexes is challenging, as we need to identify the domains that are useful for such a model. Furthermore, it is yet unclear which of the various dynamical models and coupling approaches will be most suitable to describe macromolecular dynamics. In contrast to simple physics models such as Ising models, there is no a priori definition of discrete sites, spins or subdomains in a protein system, nor is the number or discreteness of states within each subsystem well-defined. Their definition will instead largely depend on the observables in which one is interested in computing. Here, we therefore introduce Markov Field Models (MFMs) as an umbrella term and discuss recent progress and ideas in this direction. 
A key idea is Independent Markov Decomposition (IMD), which is an approach that spatially decomposes a system into independent domains that are subsequently described by domain-specific (uncoupled) Markov State Models.
In this review, we trace the path from IMD to higher models in which the domains are coupled, such as Dynamical Graphical Models. We discuss a) how domain decompositions can be estimated from data and b) ideas to represent the global thermodynamics and kinetics in terms of a coupled local dynamics between such domains.

\section*{Markovian dynamics}

Markovian models describe the dynamics generated by an operator $\mathcal{P}$, which propagates the probability density $\rho$ of a system over a finite time $\tau$ given an initial probability density $\rho_0$ \cite{prinz_markov_2011}, 
\begin{equation*}
    \rho_\tau = \mathcal{P}(\tau) \circ \rho_0.
\end{equation*}
These dynamics are usually approximated by lumping protein conformations into $M$ discrete global states. In this case, the operator $\mathcal{P}(\tau)$ becomes the transition matrix $\mathbf{P}(\tau)$ with its element $(i, j)$ describing the conditional probability to jump from state $i$ to state $j$ within a lag time $\tau$. Furthermore, the probability densities $\rho$ become vectors $\mathbf{p}$ that encode the probability distribution over the discrete states. The dynamics of the resulting Markov State Model (MSM) is then
\begin{equation}
	\mathbf{p}^T_\tau = \mathbf{p}^T_0 \mathbf{P}(\tau),
\end{equation}
with $(\cdot)^T$ denoting the vector transpose. This model describes the kinetics of the underlying system and can, e.g., be depicted as a network of exchanging states as in Fig.~\ref{fig:graph}a.
In this review, we limit our discussion of Markovian models to discrete state MSMs, while a whole zoo of such models exist, including transfer operator and Koopman operator models, Master-equation models and fuzzy MSMs. Please compare Ref.~\cite{noe_machine_2020} for references and the DeepTime library \cite{hoffmann_deeptime_2022} for software implementations.

\begin{figure}
	\centering
	\includegraphics[width=\columnwidth]{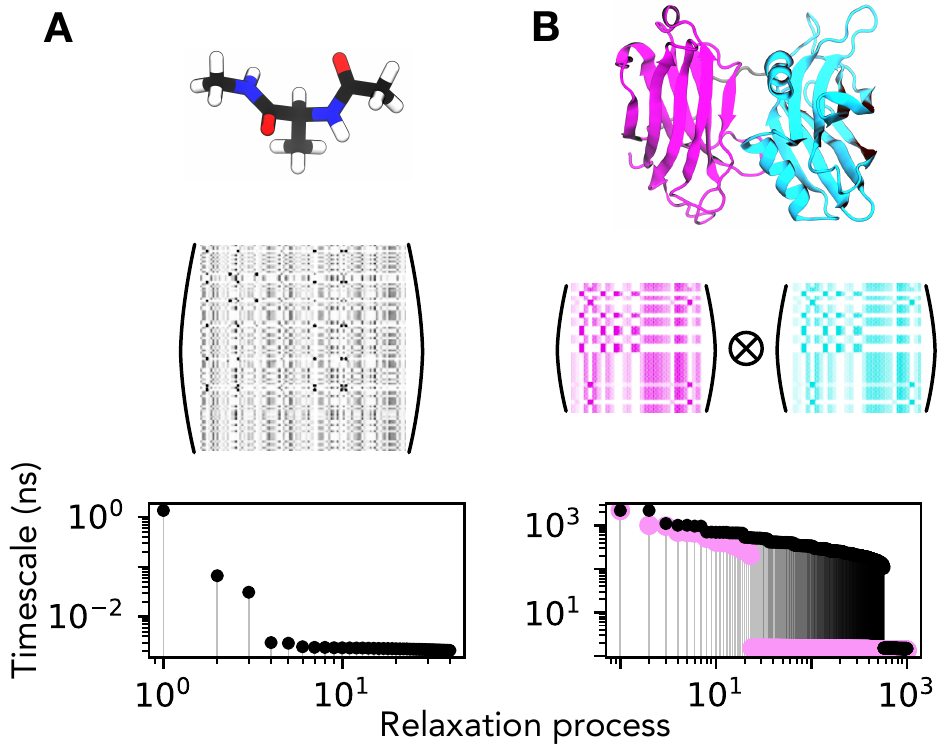}
	\caption{\textbf{Transition matrices and implied timescale spectra of a simple test system and a multi-domain protein.} Descriptions from top to bottom. \textbf{a:} Capped alanine, depicted in licorice representation, as represented by 100x100 transition matrix. The spectrum shows three implied timescales that describe slow processes. \textbf{b:} Syt-C2AB consists of a C2A and a C2B sub-domain \cite{sudhof_calcium_2012,fuson_structure_2007}, color-coded magenta and cyan. Assuming equal dynamics in both domains and no coupling, %
	the combined spectrum has > 550 slow implied timescales (black dots). We note that this represents an upper bound as the real number may be reduced by domain-domain couplings. For comparison, the spectrum of only the C2A domain has already 23 slow processes above the implied timescales gap (magenta dots), as identified by our previous work \cite{hempel_independent_2021b}.}
	\label{fig:spectra}
\end{figure}

We usually assume that the eigenvalue spectrum of the MSM transition matrix has a small number of eigenvalues close to 1 at a given lagtime $\tau$, corresponding to only a few number of transition processes whose autocorrelation times significantly exceed $\tau$, and therefore are slow relative to $\tau$.
This is a mild assumption which has been found to be practically useful for many small- to mid-sized proteins which have folded structures and are therefore sufficiently cooperative.
For example, a popular atomistic model of capped alanine has three slow processes on the nanoseconds timescale which may be identified by finding the number of implied timescales above the implied timescales gap (Fig.~\ref{fig:spectra}a, \cite{prinz_markov_2011}). 
In our atomistic model of the C2A domain of Synaptotagmin, we find 24 metastable states connected by processes on the 100s of nanoseconds timescale \cite{hempel_independent_2021b}, whereas we expect this number to be in the 100s for the C2AB dimer (Fig.~\ref{fig:spectra}b). Parametrizing a global MSM for this system would require extensive MD simulations, likely on the order of multiple milliseconds in aggregate trajectory data.

\section*{Markov field models}
The abstract idea of MFMs is to model the global transition matrix $\mathbf{P}$ by a tensor product of smaller transition matrices $\mathbf{P}_i$ that govern the dynamics of $N$ local domains, and a coupling term $\mathbf{Y}$ which models the statistical dependence between individual Markov domains  \cite{gelss_tensortrain_2017,gelss_nearestneighbor_2017}: 
\begin{equation}
	\mathbf{P} \approx \bigotimes_{i=1}^N \mathbf{P}_i + \mathbf{Y}.
	\label{eq:general_decomposition}
\end{equation}
In Fig.~\ref{fig:comparison}, we compare two methods that either explicitly model coupling terms $\mathbf{Y}$ or discard them alltogether.
Intuitively, MFMs are suitable for larger systems as the number of model parameters stored on the right side of Eq.~\ref{eq:general_decomposition} may be smaller compared to the left side, saving memory and increasing statistical efficiency as the number of independent parameters that have to be estimated is smaller.

As indicated in Eq.~\ref{eq:general_decomposition}, MFMs can in principle generate the global transition matrix $\mathbf{P}$ as a function of local transition matrices $\mathbf{P}_{i}$ and their coupling, and they therefore allow us to compute thermodynamic and kinetic quantities and compare to experimental data. However, in contrast to MSMs, it is not necessary and often not feasible to actually compute $\mathbf{P}$ explicitly. As an example consider an Ising model with $N$ spins which possesses $2^{N}$ global states with a $2^{N}\times2^{N}$ transition matrix. Instead of computing and analyzing $\mathbf{P}$ directly, we will thus usually sample the dynamical models by running a simulation algorithm that employs $\mathbf{P}_{i}$ and $\mathbf{Y}$, e.g., some form of Markov-Chain Monte Carlo.

Compared to global-state models, MFMs come with the challenge of finding a meaningful decomposition of the molecular system into domains (nodes) and learning the interaction graph (edges), which is a problem that has been extensively studied \cite{parise_structure_2006, olsson_dynamic_2019}.
In the following, we will review methods that seek to approximate the right-hand side of Eq.~ \ref{eq:general_decomposition}.

\section*{Independent Markov decomposition}
\label{sec:imd}
\begin{figure}
    \centering
    \includegraphics[width=0.8\columnwidth]{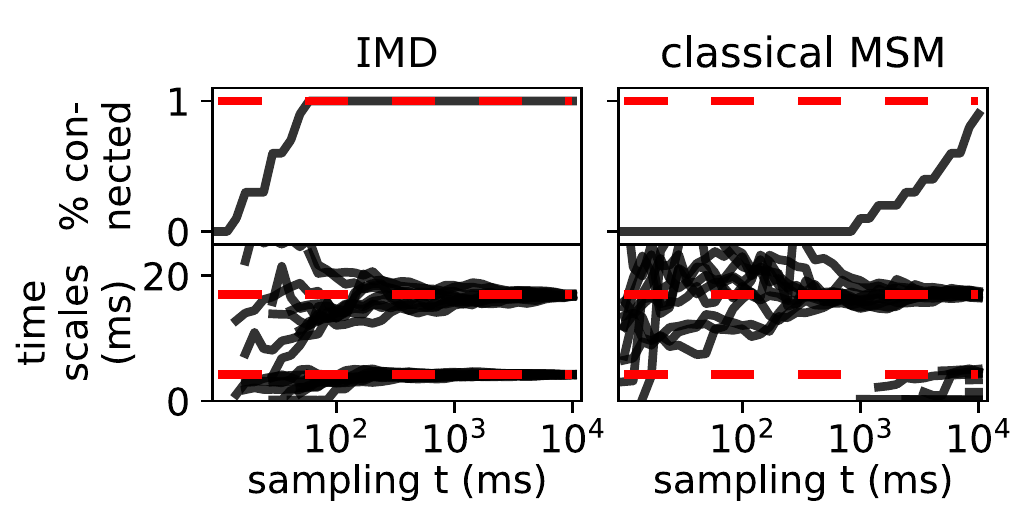}
    \caption{\textbf{Statistical efficiency of IMD versus MSM.} How much sampling is necessary for reconstructing the Hodgkin–Huxley model from simple discrete data? IMD (left column) is compared with classical MSMs (right column) for ten realizations of the underlying random process. The percentage of connected (valid) transition matrices in this ensemble is assessed as well as representative implied timescales (ground truth given by red dashed lines). Figure reproduced from Ref.~\cite{hempel_independent_2021b}.}
    \label{fig:imd_sampling}
\end{figure}

When the local domain Markov processes are uncoupled, i.e., statistically independent of each other, Eq.~\ref{eq:general_decomposition} simplifies to the Kronecker product of local transition matrices, a model termed Independent Markov Decomposition (IMD) \cite{hempel_independent_2021b},
\begin{equation}
	\mathbf{P} = \bigotimes_i \mathbf{P}_i.
	\label{eq:imd}
\end{equation}
An IMD model approximates the system dynamics by neglecting interaction terms between domains, i.e., does not model couplings (the edges in Fig. \ref{fig:graph}d are assumed to be negligible to first order).
Even though this assumption is not expected to completely model the global kinetics, its strength is that it can reduce the simulation data required to get sufficient sampling by orders of magnitude compared to an MSM (Fig.~\ref{fig:imd_sampling}) and it extracts approximately independent Markov domains from data, often leading to a much more straightforwardly interpretable model compared to a global MSM (Tab.~\ref{tab:classification}). Higher-order coupling terms $\mathbf{Y}$ can then be added a posteriori in order to better approximate the global dynamics.

Based on the Variational Approach to Markov processes (VAMP) \cite{wu_variational_2019}, an independence score could be defined that allows to partition a system into domains, maximizing the metastability in the state definition of each domain.
IMD has been shown to be applicable to high-dimensional MD data \cite{hempel_coupling_2020a}. An application example is shown in Fig.~\ref{fig:comparison}.
In a mathematical context, IMD can be considered as an estimator for the \textit{nearest Kronecker product problem} \cite{vanloan_ubiquitous_2000, loan_approximation_1993} which describes the decomposition of a matrix $\mathbf{A}$ into a product $\mathbf{B} \otimes \mathbf{C}$. 

\section*{iVAMPNets}
\begin{figure}
    \centering
    \includegraphics[width=\columnwidth]{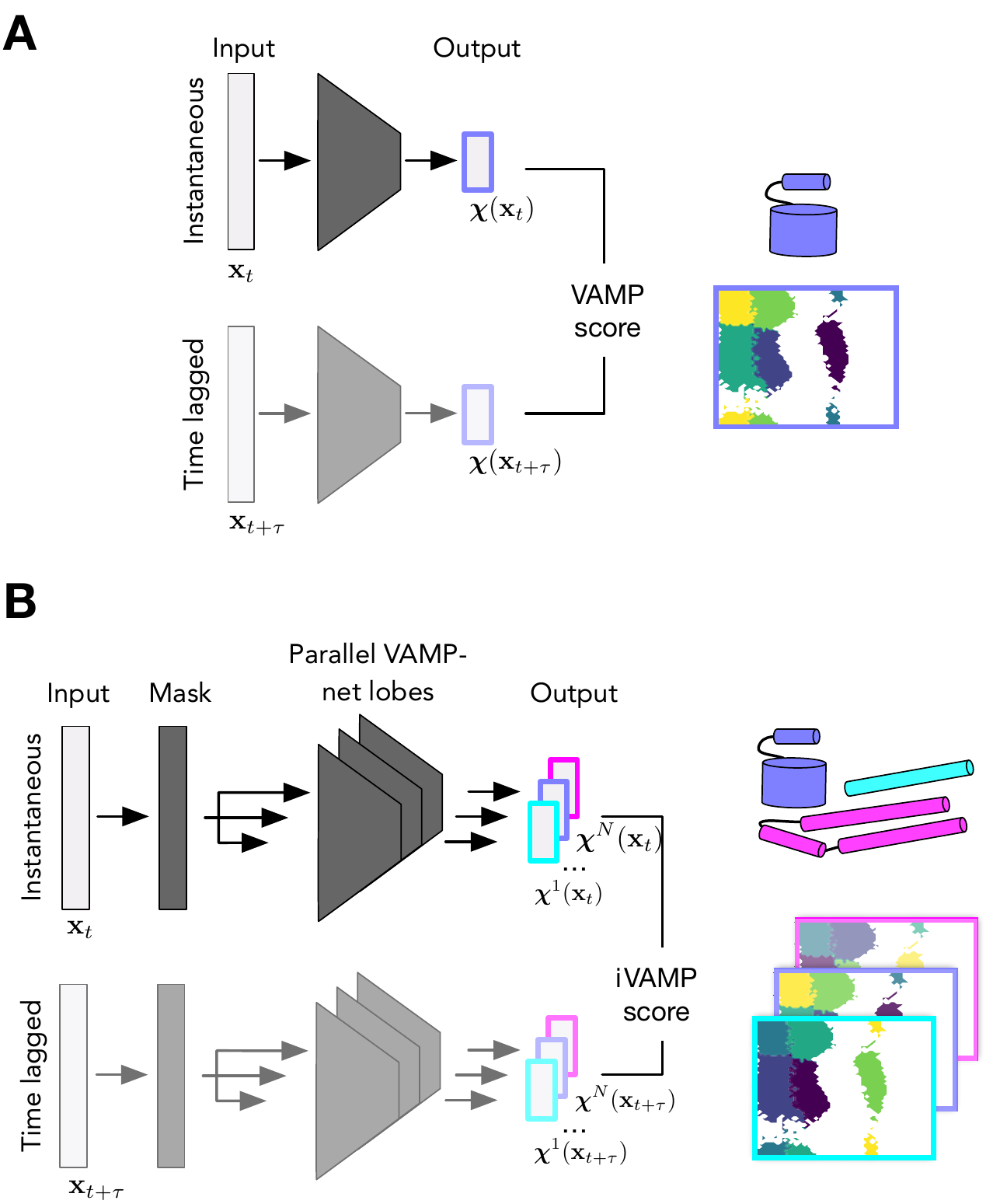}
    \caption{\textbf{Performing an independent Markov decomposition with iVAMPNets.} Simplified neural network architecture of VAMPNets (\textbf{a}) and iVAMPNets (\textbf{b}). Dark shaded shapes contain trainable weights. Domain assignments are color-coded and associated with sample projections of VAMPNet metastable assignments.}
    \label{fig:ivampnets}
\end{figure}

The idea of IMD was later combined with VAMPNets~\cite{mardt_vampnets_2018} to iVAMPNets~\cite{mardt_deep_2022}, an end-to-end unsupervised deep learning system which performs IMD given MD simulation data. To this end, iVAMPNets learn a probabilistic partitioning of the protein structure to approximately independent Markov domains by using an attention mechanism, and then uses VAMPnets in order to learn the nonlinear coordinate transform into collective coordinates resolving the rare events in these individual domains as well as a partitioning into their local Markov states. The local MSMs $\mathbf{P}_{i}$ can then be easily extracted.
iVAMPNets are trained by minimizing a loss function that  combines VAMP \cite{wu_variational_2019} and IMD \cite{hempel_independent_2021b}. The deep neural network architecture is shown and compared to the original VAMPNet architecture in Fig.~\ref{fig:ivampnets}.

\section*{Ising and Potts models, graphical models, and Markov random fields}
Arguably the most famous models that
rely on local properties instead of global state space descriptions are the Ising model \cite{ising_beitrag_1925} (Fig.~\ref{fig:graph}b) and the Potts model \cite{wu_potts_1982}. They model equilibrium distributions of coupled local domains, which are called spins due to their original use-case of modeling magnetic materials.
Ising and Potts models were successfully applied to biological systems, e.g., in protein folding \cite{munoz_what_2001,henry_comparing_2013}) or direct-coupling analysis (DCA) \cite{weigt_identification_2009}. 
The Ising model was extended to a continuous-time dynamical model by Glauber \cite{glauber_timedependent_1963,ito_glauber_1997}. The topology of the model is defined by means of a coupling graph (Tab.~\ref{tab:classification}). For the simplest case of a linear periodic chain of spins with nearest-neighbor couplings and no external field, Glauber dynamics defines the rate of a spin $i$ to flip its state $\sigma_i \in \{-1, 1\}$ as a function of its neighbors,
\begin{equation}
	q_i = q_i(\sigma_{i-1}, \sigma_{i+1}) = \frac{\alpha}{2}  -  \gamma \frac{\alpha}{2} \sigma_i \frac{\sigma_{i-1} + \sigma_{i+1}}{2}
	\label{eq:glauber}
\end{equation}
with the flipping rate of an independent spin $\alpha/2$ and the coupling parameter $\gamma$. Sampling spin-flips from these rates (e.g. using a Markov chain sampler) will sample from the Boltzmann distribution of the Ising model Hamiltonian and define dynamics in the sense of an MFM. To compare with the general idea of MFMs (Eq.~\ref{eq:general_decomposition}), we can interpret the $\alpha/2$ term as the transition rate ignoring the coupling with neighbor spins and the second term as the coupling term.

Ising model approaches can be generalized to higher dimensions, other coupling graphs, other spin Hamiltonians with external fields, etc. When coupling graphs are used that do not correspond to a regular lattice topology, one usually speaks of the more general Markov Random Fields (MRF) or graphical models. Both models have been extensively used in statistics and machine learning in order to express the conditional dependence of random variables. In MRFs, which are a generalization of Ising and Potts models, this dependence is expressed by an undirected graph – for example the probability distribution of spin $\sigma_{i+1}$ in the 1D Ising model conditionally depends on $\sigma_{i}$ and vice versa. The classical periodic Ising chain in one dimension would be depicted as a circle with every node interacting with its left and right neighbors. MRFs can be efficiently evaluated using energy functions defined over so-called \textit{cliques} (i.e., fully connected subgraphs) \cite{preston_gibbs_1974,kindermann_markov_1980}. Conceptually, MRFs directly relate to the graph shown in Fig.~\ref{fig:graph}d.

Graphical models are another generalization of MRFs and use directed graphs, i.e., in a graphical model it is possible that the distribution of random variable $\sigma_{i+1}$ conditionally depends on $\sigma_{i}$ but not vice versa \cite{murphy_dynamic_2002}.

\section*{Dynamic graphical models}
MRFs and graphical models specify the conditional dependence of random variables and define an equilibrium distribution when sampled. However, they both do not define the dynamics, e.g., they do not model how often a given spin flip or state transition is attempted in a given physical time window. In order to model biomolecular kinetics, we therefore need to additionally specify a dynamical model.

\begin{figure}
    \centering
    \includegraphics[width=\columnwidth]{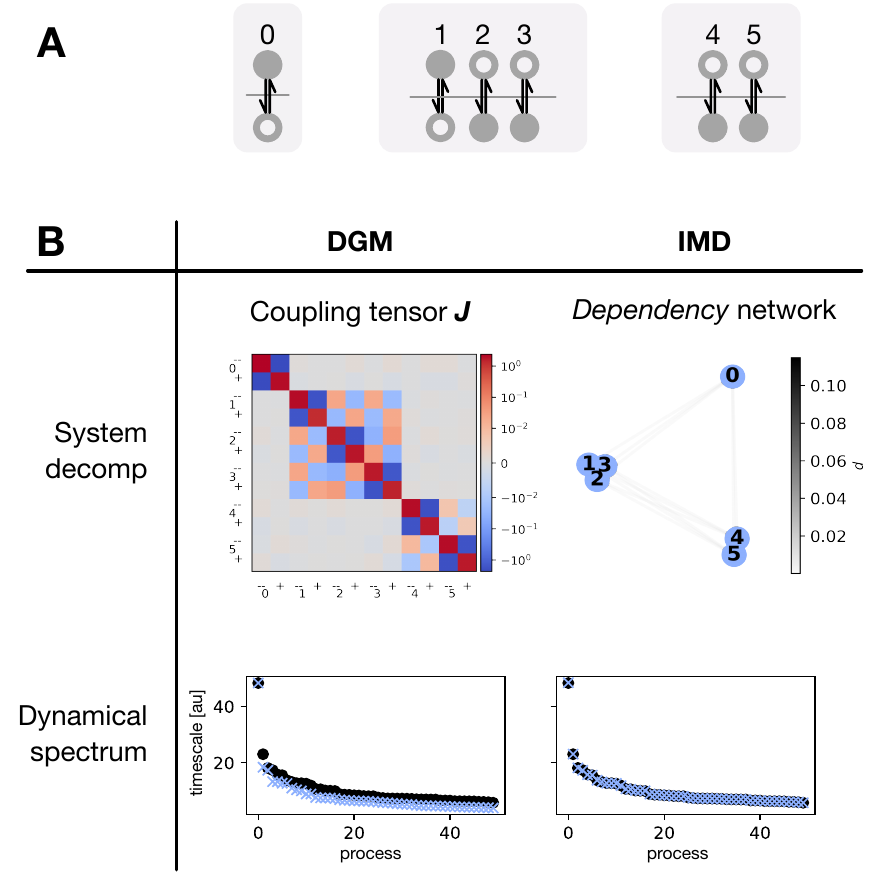}
    \caption{\textbf{Comparison of a method explicitly modeling coupling terms (DGM) and one discarding them (IMD).} \textbf{a:} The test system is a set of spins with coupling which is constructed such that spin groups $[0]$, $[1,2,3]$, and $[4, 5]$ are mutually independent. \textbf{b:} Modeling results of DGMs (left) compared to IMD (right). DGMs explicitly model the coupling in a tensor, shown as a matrix here, whereas IMD qualitatively groups coupled spins together in a graph plot. The implied timescales spectrum is well approximated by both methods, IMD matching the ground truth to numerical precision as the test systems are truly independent.}
    \label{fig:comparison}
\end{figure}

In their 2017 paper \cite{gerber_direct_2017}, Gerber \& Horenko derive a method to model a system's dynamics from domain-specific feature vectors, e.g., encoding discrete backbone dihedral states. 
Although it is not directly based on MRFs or graphical models, it can be considered to be in this tradition.
Olsson \& Noé \cite{olsson_dynamic_2019} derived a method explicitly developed as an extension of MRFs to dynamical systems under the name Dynamic Graphical Models (DGMs).
Given a set of known domains, DGMs address the problem of learning the coupling graph (Tab.~\ref{tab:classification}) and, subsequently, the global transition matrix from the data by solving an independent logistic regression problem per domain.
They assume conditional independence of the future states of a domain, given all domain states at a previous time. More specifically, DGMs model the transition probabilities for a domain configuration $\mathbf{s}_t = (\sigma_{0,t}, \sigma_{1,t}, \dots, \sigma_{N,t})$ given an initial configuration $\mathbf{s}_{t-\tau}$. The coupling is explicitly modeled with a coupling tensor $\mathbf{J} =\mathbf{J}(\tau)$. To compare to the general idea (Eq.~\ref{eq:general_decomposition}), we re-write the DGM transition probabilities in the binary case and in absence of an external field, 
\begin{equation}
\begin{gathered}
p(\mathbf{s}_t| \mathbf{s}_{t-\tau}) \sim \\ \exp\left[ \sum_{i=1}^N \sigma_{i,t}^{\top}\mathbf{J}_{ii} \sigma_{i, t-\tau} + \sum_{i=1}^N \sigma_{i,t}^{\top} \sum_{\substack{j=1\\ j \ne i}}^N \mathbf{J}_{ij}\sigma_{j,t-\tau}\right],\label{eq:olsson}
\end{gathered}
\end{equation}
highlighting how self- and pair-coupling terms enter the model. Here, $\sigma_{i,t}$ encode the $i$'th domain state at time $t$ and $\mathbf{J}_{ij}$ is a sub-matrix of $\mathbf{J}$ encoding the coupling between domains $i$ and $j$.
We note that Eq.~\ref{eq:olsson} yields a probability for a given global spin configuration $\mathbf{s}$; the construction of a global transition matrix is equivalent to the Glauber model.
An example of a DGM application and the coupling tensor is shown in Fig.~\ref{fig:comparison}.

\section*{Stochastic automata networks}
Stochastic automata networks (SANs) \cite{plateau_stochastic_1985, plateau_stochastic_2000} are MFMs that aim to reduce the dynamics of a system with a large state space into a network of weakly coupled random processes (or stochastic automata) using Kronecker products.
They operate on a known set of domains and model couplings with fixed functional forms (Tab.~\ref{tab:classification}), which are occasional synchronization events and environment-dependent changes of transition probabilities.
SANs were originally developed in the computer science community and have been applied to biophysical applications, for example, for modeling the state transition dynamics of coupled Ca${}^{2+}$ channels \cite{nguyen_stochastic_2005}.

Within the SAN formulation, the transition matrix of the global system $\mathbf{P}$ can be constructed from the local domain transition matrices $\mathbf{P}_i^{(l)}$ and matrices $\mathbf{P}_i^{(s)}$ and $\mathbf{P}_{i}^{(s, n)} = \text{diag}(\sum_{\text{rows}}\mathbf{P}_i^{(s)})$ that encode synchronization events\footnote{Simplified notation, please consult \fullcite{plateau_stochastic_1991} for details.} $s \in \epsilon$.
The lower index $i$ denotes the local system, $\epsilon$ is the set of synchronization events, and $l$ is the label that determines the type of transition. 
The global transition matrix can be written as \cite{plateau_stochastic_1991}
\begin{equation}
	\mathbf{P} = \bigotimes_{i=1}^N \mathbf{P}_i^{(l)} + \left[ \sum_{s\in\epsilon} \bigotimes_{i=1}^N \mathbf{P}_i^{(s)} - \bigotimes_{i=1}^N \mathbf{P}_{i}^{(s, n)} \right],
	\label{eq:san}
\end{equation}
which again can be interpreted in light of the general idea of MFMs (Eq.~\ref{eq:general_decomposition}).
When stochastic automata --  or, in other words, Markov processes -- are mutually independent, the SAN equation (Eq.~\ref{eq:san}) simplifies to the Kronecker product of the local transition matrices \cite{plateau_stochastic_2000} which is equivalent to IMD (Eq.~\ref{eq:imd}).

\begin{table}
    \begin{tabular}{l|ll}
    \textbf{Model}       & \textbf{Domains} & \textbf{Couplings}    \\ \hline
    \textbf{Glauber dynamics} & pre-specified        & pre-specified       \\
    \textbf{DGM}         & pre-specified        & learned      \\
    \textbf{IMD}         & learned               & n/a \\
    \textbf{iVAMPNets}   & learned               & n/a \\
    \textbf{SAN}         & pre-specified        & pre-specified      \\
    \end{tabular}
    \caption{\textbf{Classification of different Markov field models.}}
    \label{tab:classification}
\end{table}

\section*{Related models}
A slightly different approach to modeling dynamical systems which may be seen as related to MFMs is causality modeling, which is
often based on Granger's notion of causality \cite{granger_investigating_1969} or Schreiber's definition of transfer entropy \cite{schreiber_measuring_2000}.
Causality models often estimate directed graphs between local domains but without attempting to model the global dynamics of the system. Causality modeling may, e.g., proceed by inferring pairwise causality between measurement channels of recorded time-series data \cite{quinn_estimating_2011}.

Gerber and Horenko present a mathematically rigorous approach to estimate such models for MD simulations \cite{gerber_inference_2014}.
Estimating the causality graph from time-series data of short peptide dihedral torsion angles, they shed light on the spatial and temporal structure of residue-residue interactions.
Furthermore, causality modeling has been applied to quantify allosteric couplings in proteins \cite{hacisuleyman_entropy_2017,hempel_coupling_2020a}, i.e., to model directional influence between spatially distant residues or loops.

\section*{Discussion and Outlook}
MFMs take the view that the dynamics of a complex system, e.g., a protein, can be modeled by multiple weakly coupled, or independent dynamic domains.
Even though MFMs are proper generalizations of global-state methods such as MSMs, they only unfold their full potential if subsystems exist such that their coupling is weak or negligible, or when the subsystems and their coupling share similarities. In these cases MFMs will require fewer parameters than MSMs and other global dynamic models in which each transition probability is assumed to be an independent parameter, and MFMs are then more statistically data efficient , avoiding the exponential scaling problem of MSMs for large systems. Even in cases where MFMs are not statistically more efficient, learning MFMs from data can be very insighful as they can partition large dynamical systems into smaller, weakly coupled domains and present a simpler and better interpretable picture of the individual domain dynamics and their coupling than a global state model in whose parameters the local dynamics and their couplings are compounded.

The MFM concept has already been successfully applied as a simulation model to generate data on large molecular complexes. Notable examples are Ultra-Coarse-Graining (UCG), which treats local domains as multi-state coarse-grained beads \cite{dama_theory_2013, grime_coarsegrained_2016} (compare Fig.~\ref{fig:graph}d),
or MSM/RD, which describes whole proteins as multi-state particles in a reaction-diffusion (RD) setting \cite{dibak_msm_2018,delrazo_multiscale_2021}. 
Therefore, we believe that MFMs have a broad application basis for recent and future problems of kinetic modeling.

Research on MFMs is still in its infancy and faces significant challenges. First, learning the dependency graph between subdomains is a highly nontrivial task, and a well-known hard problem in computer science \cite{wainwright_highdimensional_2006,zheng_dags_2018}. 
Algorithms must be able to cope with real world application cases that range somewhere in between the extreme cases
of strongly coupled systems and systems with well-defined, completely independent domains. 
Finding the right trade-off between these extremes is related to defining the domain decomposition optimally. Compared to the hard problem of identifying the true graphical model structure underlying a dataset, MFMs for simulation data are more forgiving in that we are usually mainly interested in learning an MFM that can correctly model statistical observables of the overall systems. This task can be completed, within statistical uncertainty, even when it is not possible to uniquely determine a single MFM structure from the simulation data.

Second, systematic model errors are introduced by neglecting or sparsifying couplings, possibly both at the local domain level and globally.
Therefore, it is important to quantitatively validate MFMs, to model couplings explicitly with methods such as DGMs \cite{olsson_dynamic_2019} and / or to merge strongly coupled subsystems in such analyses.
Quantifying and controlling such model errors is a task for future work. 

Third, validating computational models with experiments usually requires us to estimate the ensemble averages of the global system.
In MFMs, it is often unfeasible to explicitly construct and analyze the global transition matrix $\mathbf{P}$, which means we have to compute its properties by sampling the coupled subsystem dynamics, which results in statistical errors and possibly sampling problems in order to compute such ensemble averages.

In summary, we believe that MFMs have great potential for modeling and understanding the dynamics of large-scale biomolecular complexes and may significantly reduce the sampling problem, thus advancing the applicability of computationally expensive, high-resolution simulation methods such as atomistic MD. However, many challenges need to be solved until these models have reached a similar maturity and practicality as MSMs, providing fertile soil for future investigations.

\section*{Acknowledgements}
TH thanks Patrick Gel\ss\ for fruitful discussions. We acknowledge the financial support of Deutsche Forschungsgemeinschaft DFG (SFB/TRR 186, Project A12), the European Commission (ERC CoG 772230 \textquotedblleft ScaleCell\textquotedblright), the Berlin mathematics research center MATH+ (AA1-6), and Berlin Institute for the Foundations of Learning and Data (BIFOLD). This work was partially supported by the Wallenberg AI, Autonomous Systems and Software Program (WASP) funded by the Knut and Alice Wallenberg Foundation (to S.O.).

\printbibliography
\addcontentsline{toc}{section}{References}

\end{document}